\documentclass{aastex}
\usepackage{emulateapj5}
\usepackage{onecolfloatx}
\begin{document}

\twocolumn[

\def\tm{$\m@th^{^{\rm T\kern -.04em M}}$}

\def\et{{\it et al.}}
\def\cbr{\mbox{\scriptsize CMB}}
\def\gap{\stackrel{>}{_\sim}}
\def\lap{\stackrel{<}{_\sim}}
\def\TBD{\bf TBD}

\title{A Limit on the Polarized Anisotropy of the Cosmic
Microwave Background at Subdegree Angular Scales}
\author{M. M. Hedman, D. Barkats, J. O. Gundersen\altaffilmark{1}, 
	S.T. Staggs\altaffilmark{2}, B. Winstein\altaffilmark{3} }

\affil{Joseph Henry Laboratories and Physics
Department,\\ Princeton University, Princeton, NJ 08544}

\begin{abstract}

A ground-based polarimeter, PIQUE, operating at 90 GHz has set a
new limit on the magnitude of any polarized anisotropy in the
cosmic microwave background.  The combination of the scan
strategy and full width half maximum beam of $0\fdg 235$ gives
broad window functions with   $\langle\ell_E\rangle =
211^{+294}_{-146}$ and $\langle\ell_B\rangle = 212^{+229}_{-135}$
for the E- and B-mode window functions, respectively.  A joint
likelihood analysis yields simultaneous 95\% confidence level
flat band power limits of 14 and 13 $\mu$K on the amplitudes of
the E- and B-mode angular power spectra, respectively. Assuming no
B-modes, a 95\% confidence limit of  10 $\mu$K is placed on the
amplitude of the E-mode angular power spectrum alone. 
\end{abstract}

]

\altaffiltext{1}{Robert H. Dicke Fellow}
\altaffiltext{2}{Alfred P. Sloan Fellow}
\altaffiltext{3}{Department of Physics and the Enrico Fermi Institute,
University of Chicago, Chicago, IL 60637.}

\section{INTRODUCTION} 
\nopagebreak 
Observations of the temperature anisotropies of the cosmic microwave background (CMB) already constrain cosmological models (e.g. Tegmark \& Zaldarriaga 2000)\nocite{tegzal00}.  In addition,
a linearly polarized component of the CMB arises from any quadrupolar variation in the photons
scattered from electrons at the  last scattering surface \cite{ree68,huwhi97}.
Polarization data will complement other CMB data (e.g., Kosowsky 1999).
Current cosmological models predict that polarization anisotropies are 10--20 times smaller than temperature anisotropies.
To date only upper limits on the CMB polarization exist, summarized in Staggs et al. (1999)\nocite{sta99} and Subrahmanyan et al. (2000)\nocite{ATCA}. 
The polarization fluctuations may be decomposed into gradient
and curl parts \cite{kamkow97}, here designated E- and B-modes 
\cite{zalsel97}.  The complete formalism for analyzing
polarization data developed by these
authors, and extended in Zaldarriaga, 1998\nocite{zal98}
(henceforth, Z98),
is applied to data here for the first time.
  
\section{INSTRUMENT}
\nopagebreak

The Princeton IQU Experiment\footnote{Three of the four Stokes
parameters are I, Q, and U.  PIQUE measured I and Q in 2000 and
will add U in 2001.} (PIQUE) comprises a single 90~GHz 
correlation polarimeter
underilluminating a 1.2~m off-axis parabola \cite{wol97}.  
The beam on the sky
is $0\fdg 235$, and the instrument observes a single Stokes
parameter in a ring of radius $1^\circ$ around the NCP.  The
receiver is a heterodyne analog correlation polarimeter   using
W-band (84--100~GHz) HEMT amplifiers.   A
mechanical refrigerator cools the corrugated feed horn, the
orthomode transducer (OMT), and the
HEMT amplifiers to $\lesssim 40$~K.  The telescope is fixed in
elevation, but rotates in azimuth.
Two large nested ground shields surround the instrument; the inner
shield rotates with the telescope.

The correlation polarimeter \cite{krauss} directly measures the
polarized electric field, rather than detecting and then differencing two
large intensity signals.   The two input signals come from an OMT oriented
so one arm is parallel to  the azimuthal scan direction.  The
polarimeter output is proportional to the linear polarization for
axes rotated $45^\circ$ with respect to the inputs.   The local
oscillator signal in one polarimeter arm is phase-switched at
4~kHz, well above  the 1~kHz $1/f$ knee of the amplifiers
\cite{wolpos98}.   The 2--18~GHz intermediate frequency (IF) band is split into three
frequency channels (S0, S1, S2).   The total power in each of the
IF arms is also detected, though with significantly less sensitivity
since the phase-switching does not apply.  The polarimeter characteristics
are given in Table~\ref{ta:receiver}.

\begin{deluxetable}{clccc}
\tablecaption{PIQUE 2000 Polarimeter Characteristics\label{ta:receiver}} 
\tablewidth{252pt} 
\tablehead{ 
& \colhead{$\nu_c(\Delta \nu)$\tablenotemark{a}} & \colhead
{$S$\tablenotemark{b}} & 
\colhead{$T_{off}$\tablenotemark{c}} &
\colhead{$\overline{\sigma}$\tablenotemark{d}} \\
\colhead{Channel} & \colhead{(GHz)} & \colhead{(mK~s$^{1/2}$)} & 
	\colhead{($\mu$K)} & \colhead{($\mu$K)} \\ 
 } 
\startdata 
S0 & 87.0(4.0)  & 2.3  & $-20(10)$ & 25  \\
S1 & 91.5(4.8)  & 2.3  & $330(30)$ & 26  \\
S2 & 96.2(2.3) & 3.4 & $600(30)$ & 41  \\
\enddata
\tablenotetext{a}{\footnotesize Effective center frequencies and bandwidths, including gain slope and phase match degradations.}
\tablenotetext{b}{\footnotesize Raw sensitivity measured on a clear day with $T_{atm} \approx 45$~K, uncorrected for atmospheric opacity.  The differencing strategy degrades the sensitivity by a factor of two.}
\tablenotetext{c}{\footnotesize Average chopped offset where the error
gives the standard deviation of the mean of the eleven
offsets calculated  during the season; see text.  The typical 
statistical error in calculation of 
a single one of the 11 offsets is $20~\mu$K.}
\tablenotetext{d}{\footnotesize The 
average error per bin for the final data binned into
twenty-four $0\fdg 26$-wide RA bins.  Note that the analysis
uses data in 144 bins.}
\tablecomments{\footnotesize All temperatures are in thermodynamic units.}
\end{deluxetable}

\section{OBSERVATIONS AND CALIBRATION.}

Between 2000 January 19 and 2000 April 2, PIQUE collected 
slightly over 800 hours of data from the roof of the physics building at
Princeton, latitude $40\fdg 345$, east 
longitude $-74\fdg 647$.  
Every five seconds, 
the telescope alternated between two azimuth positions, $\pm
0\fdg 93$, at elevation $41\fdg04$, measuring $\mp Q$ (as defined
by the IAU) for two regions 
separated by six hours in right ascension (RA)
on the $\delta = 89^\circ$ ring.

Since the polarimetry channels have a small ($< 23$~dB)
sensitivity to total power, PIQUE differences data taken in the east position from data taken in the west postion five seconds
later.  Residual offsets in these ``chopped" data, listed in
Table~\ref{ta:receiver}, are small and stable 
($1/f$ knee undetectable at $10~\mu$Hz in clear weather).  The offsets have 
been traced to polarized emissive pickup from the fixed ground
screen, and are removed as described below.
Since $Q$ is measured in the west and $-Q$ is
measured in the east, the chopped data comprise {\it sums}
of Stokes parameter $Q$.  

Constant elevation scans of Jupiter are used to  
determine pointing accuracy, map the beams,  
and calibrate the total power channels.
The absolute errors in elevation and
azimuth are $0\fdg 01$ and $0\fdg 02$, respectively.
The beam FWHM are $0\fdg 233(9)$ (cross-elevation) and
$0\fdg 240(14)$ (elevation), in agreement with calculations and
near-field data.  
For the purpose
of the likelihood analysis, the beam dispersions are taken as 
$\sigma = \theta^{\mbox{\tiny FWHM}}_{avg}/\sqrt{8 \ln 2} =0\fdg 1$.  
Uncertainty in the brightness temperature of Jupiter contributes to a
10\% error in the calibration of the total power channels (which
are only used to correct for the slowly varying atmospheric 
opacity during the CMB polarization observations.)

The polarimetry channels are  calibrated by observations of the
polarized emission (and reflection) from a large
ambient-temperature aluminum flat. Nutating the flat
about a vertical axis results in a peak-to-peak polarized signal
of about 30~mK. These measurements are
consistent with polarized cold load tests. 
The final calibration error is 10\%,
dominated by uncertainty in the surface resistivity
($4.0~\mu\Omega\cdot$cm) of the nutating flat.

\begin{figure}[ht]
\epsscale{0.9}
\plotone{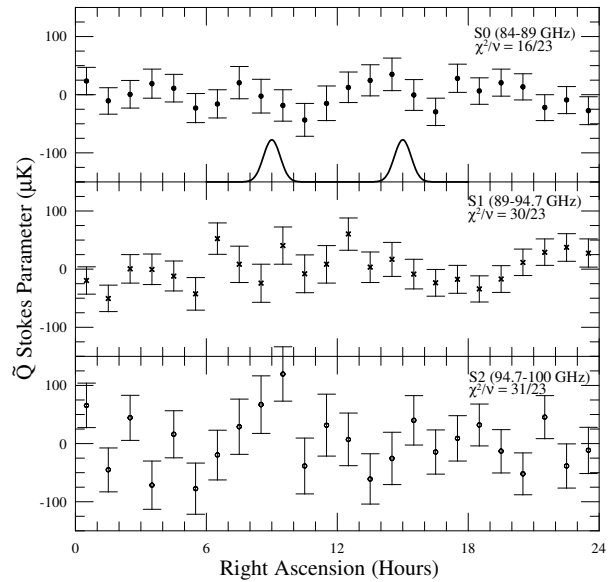}
\epsscale{0.9}
\caption{\small Binned data in thermodynamic units
for each frequency channel, versus RA. For this plot, 24
approximately beam-sized bins are shown, though the analysis
uses 144 bins.  The theoretical point source response 
to a 250~mJy polarized source is indicated in the top panel.}  
\label{fig:datalike}
\end{figure}

\section {DATA REDUCTION} 
\nopagebreak

Of the 807 hours of data, 200 hours were spent slewing the telescope, 123 hours were corrupted by various electromechanical
failures, and 52 hours were isolated segments less than eight hours in duration which are not used.  High levels of atmospheric noise contaminated 102 of the remaining 432 hours, as determined by the large positive tails in the distributions of the polarization channels' correlation
coefficients.  In order to avoid biasing the data set,
the primary atmosphere cut employs the quadrature chop data.  A quadrature chop has the raw data from
each azimuthal position (east and west) split in two to obtain 
$(E_1+W_1 -E_2-W_2)/2$, which contains no polarization
signal.  This cut removes 78 hours by requiring the absolute value of all three correlation coefficients for 
ten-minute sections of the quadrature data to 
lie below a threshold, $\eta_q = 0.32$.  Varying $\eta_q$ by $\pm5$\% does
not affect the final results statistically.  The secondary atmosphere cut culls extreme values ($>4\sigma$) of the 
CMB correlation coefficients and removes an additional 24 hours of data.

The final data cut makes use of the six hour null test.  The 
six hour null test takes advantage of the fact that certain linear 
combinations of 
data separated by six hours in RA should have zero signal: 
$d_t - d_{t+6} + d_{t+12} - d_{t+18}=0$, where $t$ is measured in hours.  
Those days that fail this null test ($p(\chi^2|\nu)<0.05$) are 
removed.  This final cut removes an additional 82 hours of data, 
leaving 248 hours of CMB data for the final analysis.  Note that 
while this cut only slightly increases the final limits, it 
significantly improves the H1-H2 null test
described below.

The remaining time series is binned into $\approx 2100$
10-minute-wide bins.  The  means,
standard deviations and inter-channel correlations are
calculated for each bin.  The attenuation due to atmospheric
opacity for each bin is estimated from 
a total power channel and the appropriate ($\approx
15\%$) correction is applied. The data are divided into eleven
periods separated from one another by stretches of weather
severe enough to require covering the instrument. The periods have
varying lengths $t_i$, where $16~\mbox{hr} < t_i < 110~\mbox{hr}$.  
An offset for each channel is calculated and removed from each period.
Other divisions of data, in which as few as two or as many as 24
offsets are removed, do not change the limit by more than $1~\mu$K.
The weighted means, standard deviations, and covariances are
calculated for 144 10-minute RA bins.  These data are rebinned for presentation in Figure~\ref{fig:datalike}.

\begin{deluxetable}{lrrrr}
\tablecaption{Results of $\chi^2$ consistency tests. \label{ta:chisq}} 
\tablewidth{252pt} 
\tablehead{ 
\colhead{DATA\tablenotemark{a}} & \colhead {$\nu$\tablenotemark{b}} 
& \colhead{S0\tablenotemark{c}} &
\colhead{S1\tablenotemark{c}} & \colhead{S2\tablenotemark{c}} \\
 } 
\startdata 
CMB & 143 & 0.73 & 0.11 & 0.88  \\ 
Quadrature  & 143 & 0.95 & 0.06 & 0.57  \\ 
H1-H2 & 143 & 0.28 & 0.24 & 0.37  \\ 
6-hr null & 36 & 0.19 & 0.32 & 0.97  \\ 
 & & \colhead{S0-S1} &\colhead{S0-S2} & \colhead{S1-S2}  \\ 
Inter-channel & 143 & 0.32 & 0.66 & 0.61\\ 
\enddata 

\tablenotetext{a}{\footnotesize The data sets are described in the text.}
\tablenotetext{b}{\footnotesize The number of degrees of freedom for each data set.}
\tablenotetext{c}{\footnotesize The probability of exceeding the $\chi^2$ for a given frequency channel.}
\end{deluxetable}

Table~\ref{ta:chisq} presents the results of a series of null
tests.  The null data
sets include the quadrature data,
data differenced between the first and second halves of the
observing run (H1-H2), data differences between signal channels,
and the six-hour null data.  The $\chi^2$ distribution of the null tests is 
consistent with noise. Similarly, no contamination is found when
the data are binned in Sun-centered or Moon-centered coordinates.

\begin{figure}
\epsscale{0.9}
\plotone{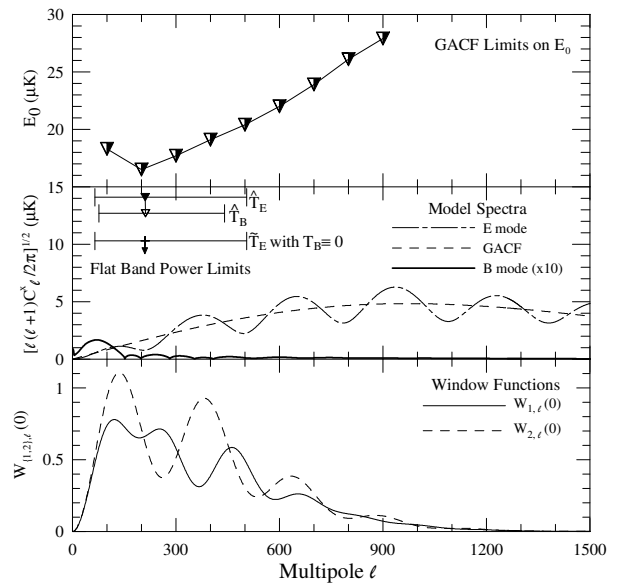}
\caption{\small Zero-lag window functions for E- and
B-modes (solid and dashed lines) are plotted in the lower
panel.  The middle panel shows the limits on $T_E$ and $T_B$
described in the text, plotted over E-mode 
predictions (dash-dot) from the
$\Omega_{tot}=1$ model of Jaffe et al. (2000) and B-mode (solid, assuming $T/S=1$). The smooth
dashed curve is
a GACF with $E_0=5.6~\mu$K and $\ell_c=687$.
To
illustrate PIQUE's ability to limit the fluctuations where they
are thought to be sizeable, the top panel shows
the 95\% confidence limits on the amplitude of the GACF, $E_0$,
where the abscissa is $\ell_{p} = \sqrt{2}\ell_c$.}
\label{fig:jogmega}
\end{figure}

\section{DATA ANALYSIS}
\nopagebreak

The Bayesian method with uniform prior is used to establish
limits on the amplitude of the polarization anisotropy and to
verify null tests of the data.  The data vector $d$ has $3N$
entries, where $N=144$.  The data covariance matrix $C_D$
consists of nine $N\times N$ diagonal submatrices, since 
correlations between 10-minute-wide bins are small enough to
neglect ($\ll 1\%$ of the variances).  Six of the
submatrices account for inter-channel correlations due to
HEMT-induced correlated gain fluctuations 
and correlations from atmospheric fluctuations weakly coupled 
into the polarimeter
channels.  The average correlation coefficients between S0-S1,
S0-S2, and S1-S2 are 0.02, 0.08 and -0.01, respectively.
The theory
covariance matrix $C_T$ may be expressed in terms of the E- and
B-mode angular power spectra $C^E_\ell$ and $C^B_\ell$ by
\begin{eqnarray} \label{eq:ctdef}
 &&C_T^{ij} =  \langle\tilde{Q}(\hat{n}_i)\tilde{Q}_j(\hat{n}_j)\rangle =\\
	 && 	\frac{1}{4\pi}\sum_{\ell =2}^\infty (2\ell +1)[
			C_\ell^E W_{1,\ell} (\phi_{ij}) +
			C_\ell^B W_{2,\ell}(\phi_{ij})],\nonumber
\end{eqnarray} 
where $\hat{n}_{i,j}$ are position vectors,
$\phi_{ij}=\cos^{-1}(\hat{n}_i\cdot \hat{n}_j)$ is the lag,
and $W_{1,\ell}$ and $W_{2,\ell}$ are associated window
functions. 
The $\tilde{Q}$ indicates that PIQUE measures
sums of $Q$'s. Z98 presents an elegant way to describe the window
functions for a ring observation strategy, here modified to
include the PIQUE differencing:

\begin{equation}
W_{\{1,2\},\ell}(\phi_{ij})=2\sum_{m=1}^{\ell}  
F^2_{\{1,2\},\ell m}(\theta)B_{\ell m}^2\cos(m\phi_{ij}),
\end{equation}

\noindent where the factor of 2 in front accounts for the negative $m$ 
values, and the $m=0$ term is
excluded from the window functions because offsets are removed from the data.
The $F_{\{1,2\},\ell m}$ are given in Z98 in terms of
associated Legendre polynomials, $\theta = 1^\circ$ is the radius
of the ring, and
\begin{equation}
B_{\ell m}=2\cos(m\alpha/2)\frac{\sin(m\Delta\phi/2)}{(m\Delta\phi/2)}e^{-\ell(\ell+1)\sigma^2/2}
\end{equation}
where $\alpha=\pi/2$ is the chop amplitude, and $\Delta\phi=\pi/72$ is the bin size.
The window functions are depicted in 
Figure~\ref{fig:jogmega}. 
The likelihood is ${\cal L} \propto \exp(-d^TM^{-1}d/2)/\sqrt{|M|}$,
where, $M=C_T+C_D$.

\begin{figure}[t]
\epsscale{1.0}
\plotone{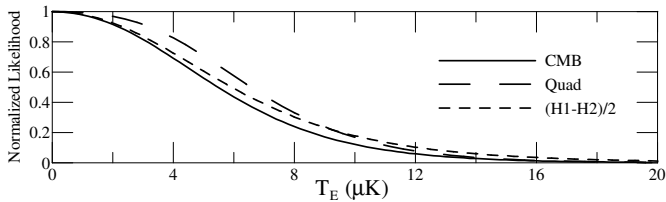}
\caption{\small Normalized likelihoods for the cases when $T_B\equiv0$.}
\label{fig:likeli}
\end{figure}

\begin{deluxetable}{ccccc}
\tablecaption{$95\%$ Confidence Limits \label{ta:likes}}
\tablewidth{252pt}
\tablehead{
\colhead{DATA\tablenotemark{a}} & \colhead{$\widetilde{T}_E$\tablenotemark{b}} & 
\colhead{$\widehat{T}_E$\tablenotemark{c}} &
\colhead{$\widehat{T}_B$\tablenotemark{c}} & 
\colhead{$E_0$\tablenotemark{d}} \\
\colhead{} & \colhead{$(\mu$K)} & \colhead{$(\mu$K)} &
\colhead{$(\mu$K)} & \colhead{$(\mu$K)} \\}
\startdata
CMB & 10 & 14 & 13  & 17 \\
Quad & 11& 14 & 13  & 18 \\
(H1-H2)/2 & 12 & 17 & 15 & 18 \\
\enddata

\tablenotetext{a}{\footnotesize The data sets are described in the text.}
\tablenotetext{b}{\footnotesize The limit $\widetilde{T}_E$ is found assuming $T_B\equiv0$.}
\tablenotetext{c}{\footnotesize The limits $\widehat{T}_E$ and $\widehat{T}_B$ are determined simultaneously 
by finding the contour of constant likelihood enclosing
95\% of the volume. }
\tablenotetext{d}{\footnotesize The limit on the amplitude of the GACF, $E_0$, at $\ell_p=200$.}
\end{deluxetable}

The likelihood analysis proceeds by first considering  flat
angular spectra, such that $\ell(\ell+1)C_\ell^X/2\pi = T_X^2$,
where $X=E, B$. Since the
amplitude of $C^B_\ell$ is predicted to be much smaller than that
of $C^E_\ell$ (see Figure \ref{fig:jogmega}), Table~\ref{ta:likes} presents the limit on 
$T_E$  under the assumption $T_B\equiv 0$.  Figure~\ref{fig:likeli} shows ${\cal L}(T_E,0)$ for the CMB data and two null data sets.  The limit $\widetilde{T}_E$ is found by
integrating ${\cal L}(T_E,0)$; the result is $\approx 30$\% higher
if ${\cal L}(T_E^2,0)$ is integrated. Next, joint
upper limits $(\widehat{T}_E, \widehat{T}_B)$ are determined by finding the
constant contour of ${\cal L}(T_E,T_B)$ enclosing 95\% of the
volume of ${\cal L}$. Finally, a gaussian autocorrelation
function (GACF) analysis is performed.
The motivation to revive the GACF is that it 
better describes the gross features of the predicted 
E-mode spectrum than the flat model; 
see Figure~\ref{fig:jogmega}.
For the GACF with characteristic scale $\ell_c$, 
the power spectra are given by
$\ell(\ell+1)C^X_\ell/2\pi = X_0^2u^2\exp(-u^2/2)$, where
$X_0=E_0,B_0$, $u\approx \ell/\ell_c$ \cite{bond95}, and $\ell_p=\sqrt2\ell_c$ is the peak of the GACF.
Only the case $B_0\equiv 0$ is considered.  
The likelihood ${\cal L}(E_0,0)$ is integrated to
place $95\%$ confidence level upper limits on $E_0$ for several
values of $\ell_p$ as shown in Figure~\ref{fig:jogmega}.
Table~\ref{ta:likes} summarizes the limits obtained for the CMB data and two null tests.
The limits in Table~\ref{ta:likes} do not include calibration
errors (10\%).  Note that for PIQUE's broad window functions, 
the 7\% beam errors only add 2\% errors in quadrature with the
calibration errors.

\section{DISCUSSION}
\nopagebreak

The formalism of Z98 enables us to place limits on both the E- and B-mode angular power spectra, allowing
direct comparison of these results with theoretical expectations. Due
to PIQUE's broad reach in $\ell$-space, these results limit
the size of the predicted acoustic peaks in the
polarization, rather than the damped tails at high and low $\ell$.
PIQUE's limits on $T_E$ and $T_B$ are higher than expected from, e.g., the
$\Omega_{tot}=1$ model of Jaffe et al. (2000).  The limits are also
higher than expected foreground levels.
Polarized dust emission anisotropy,
extrapolated from the IRAS 100 $\mu$m map \cite{IRAS}, should contribute
$\lesssim 0.5~\mu$K, while polarized synchrotron, 
extrapolated from Brouw \& Spoelstra (1976)\nocite{brouw76}
or from Haslam et al (1982)\nocite{has82}, should be less than 0.4 $\mu$K.  
Spinning dust emission is even smaller \cite{dra99}.

Since this paper represents the first effort to put limits on
$T_E$ and $T_B$ directly, it is
not possible to quantitatively compare these results to published 
results from previous experiments.  
Direct comparison would require other authors to reanalyze 
their data for sensitivity to E- and B-modes rather than to Q and U.
Qualitatively, we  note that the limit $\widetilde{T}_{E}<10$ $\mu$K
is smaller by more than a factor of two than the best limits 
previously published for $\ell < 3000$ 
\cite{torb99,Barththesis}. 
Further, PIQUE has  sensitivity to
higher $\ell$ than these two previous experiments, so this work 
represents a significant
improvement in the constraints on the polarization of the CMB.  

\acknowledgements

We are indebted to Matias Zaldarriaga for insights on analysis of
polarization experiments.  We acknowledge MZ  and Uro\v{s} 
Seljak for use of CMBFAST.  We thank Ken Ganga, Chris Herzog, 
Norman Jarosik, Lloyd Knox, Lyman Page, Paul Steinhardt, 
and David Wilkinson for many helpful discussions and
assistance. Much of the hardware for this experiment was adapted
from the Saskatoon experiments. We are grateful to Ted Griffiths
and Laszlo Varga for help in mechanical design and construction.
We also thank Marian Pospieszalski and the NRAO for supplying the
HEMT amplifers. Data (including the correlation matrix and
likelihood functions) will be made public upon publication of
this Letter.

This work was supported by a NIST precision measurement grant \#NANB8D0061
and by NSF grant \# PHY96-00015.  Additional support was provided
by the Alfred P.\ Sloan Foundation and the Guggenheim Foundation
through their Fellowships for STS and BW respectively.

\end{document}